# AN ANALYTIC MODEL FOR THE GRAVITATIONAL CLUSTERING OF DARK MATTER HALOES

H. J. Mo and S. D. M. White

Max-Planck-Institut für Astrophysik
Karl-Schwarzschild-Strasse 1
85748 Garching, Germany

# An analytic model for the gravitational clustering of dark matter haloes

**ABSTRACT**

We develop a simple analytic model for the gravitational clustering of dark haloes. The statistical properties of dark haloes are determined from the initial density field (assumed to be Gaussian) through an extension of the Press-Schechter formalism. Gravitational clustering is treated by a spherical model which describes the concentration of dark haloes in collapsing regions. We test this model against results from a variety of N-body simulations. The autocorrelation function of dark haloes in such simulations depends significantly on how haloes are identified. Our predictions agree well with results based on algorithms which break clusters into subgroups more efficiently than the standard friends-of-friends algorithm. The agreement is better than that found by assuming haloes to lie at the present positions of peaks of the linear density field. We use these techniques to study how the distribution of haloes is biased with respect to that of the mass. The initial (Lagrangian) positions of haloes identified at a given redshift and having circular velocities $v_c = v_c^*(z)$ (i.e. mass equal to the characteristic nonlinear mass $M^*$ at that redshift) are very weakly correlated with the linear density field or among themselves. As a result of dynamical evolution, however, the present-day correlations of these haloes are similar to those of the mass. Haloes with lower $v_c$ are biased toward regions with negative overdensity, while those with higher $v_c$ are biased toward regions with positive overdensity. Among the haloes identified at any given epoch, those with higher circular velocities are more strongly correlated today. Among the haloes of given circular velocity, those at higher redshifts are also more strongly clustered today. In the "standard CDM" model, haloes with $v_c = 200 \,\mathrm{km\,s^{-1}}$ and identified at redshift $z \gtrsim 2$ have present-day autocorrelation comparable to that of normal galaxies in the real universe.

# 1 INTRODUCTION

It is generally believed that the large scale structures in the universe formed through the growth of small inhomogeneities by gravitational instability. In the hierarchical clustering scenario of structure formation, a dominant dissipationless component of dark matter is assumed to aggregate into dark matter clumps, the virialized parts of which are usually called dark haloes. Galaxies then form by the cooling and condensation of gas within these dark haloes (White and Rees 1978). Non-gravitational effects, which are difficult to model, are likely to be critical in galaxy formation, yet have little effect on the formation and clustering of dark haloes. The study of the formation and clustering of haloes is therefore less ambiguous than that of galaxies, but is nevertheless important, because of the close relation between galaxies and haloes.

Dark haloes are highly nonlinear objects. Their evolution is usually studied by numerical simulations (e.g. Frenk 1991; Gelb & Bertschinger 1994a,b and references therein). Such simulations are limited both in resolution and in dynamical range and can be difficult to interpret. Our understanding of their results can be substantially enhanced by simple physical models and by analytic approximations. Also a simple and successful analytic model can be used to carry out a large parameter study without requiring a large amount of computer time. The present paper attempts to provide such a model.

The initial distribution of density fluctuations in the universe is usually assumed to be Gaussian, and so to be described completely by its power spectrum. This, in turn, is derived from a model for the origin of structure. It is perhaps feasible to associate dark haloes or galaxies with special regions of the initial density field and to consider the clustering of these objects which results both from initial conditions and from motions due to gravitational interaction. Kaiser (1984) used this idea to explain the strong clustering of Abell clusters as a consequence of the statistical properties of high peaks in an initial Gaussian field. His formalism was developed extensively by Bardeen et al. (1986, hereafter BBKS). BBKS found that if galaxies formed at high peaks, they should be more clustered than the mass, and that if galaxies of different types are associated with peaks of different heights, they should have different clustering properties. Galaxies are then biased tracers of the mass. However, it is not known how well galaxies correspond to high peaks of the initial field, and there is some direct evidence that the correspondance of such peaks with dark haloes is not particularly good (Frenk et al. 1988; Katz, Quinn & Gelb 1993). Furthermore the clustering of peaks in Eulerian space may differ substantially from that in the initial (Lagrangian) space and it is unclear how to deal with the problem that a single dark halo may contain several galaxies. Press & Schechter (1974, hereafter PS) developed a formalism which identifies haloes at a given cosmic time as regions (in the initial density field) which just collapse at that time according to a spherical infall model. Both the halo mass function they derived and the detailed structure of hierarchical clustering which their theory predicts have gained considerably in plausibility from recent theoretical work (e.g. Bond et al. 1991; Bower 1991; Lacey & Cole 1993) and from tests against N-body simulations (Efstathiou et al. 1988, hereafter EFWD; Kauffmann & White 1993; Lacey & Cole 1994). Unfortunately, these various tests show that although the statistical predictions of the theory work well, the basic hypothesis on which it is based works very poorly on an object by object basis (see Bond et al. 1991; White 1995).

The PS theory developed in the above papers does not provide a model for the spatial distribution of dark haloes. As we show below, it can, however, be extended to construct such a model. Dark haloes are defined using the initial density field as in the PS formalism. Their gravitational clustering is treated by a spherical model which describes their concentration in dense regions through their relation to the mass distribution in the initial density field. This model is tested by comparing its prediction for the two-point correlation functions of haloes with those derived from N-body simulations. We describe our model in Section 2 and compare its predictions with N-body simulations in Section 3. Section 4 discusses how spatial distributions of dark haloes of different types are biased with respect to the mass distribution. Section 5 summarizes our main conclusions. Since the analytic argument of the paper is quite conplex, we provide an Appendix which summarizes how our formulae should be used for the calculation of various correlation functions, for example the auto- and cross-correlations of haloes of differing circular velocities, and the present-day correlation properties of objects which are identified as individual haloes at high redshifts.

## 2 THE MODEL

Although the model described here may readily be extended to other cosmologies, we assume, for simplicity, an Einstein-de Sitter universe (i.e. that the total mass density parameter $\Omega = 1$, and the cosmological constant $\Lambda = 0$). All physical length scales are quoted either assuming a Hubble constant $H = 50\,\mathrm{km\,s^{-1}Mpc^{-1}}$ or in units of $h^{-1}\mathrm{Mpc}$ where $h = H/[100\,\mathrm{km\,s^{-1}Mpc^{-1}}]$.

## 2.1 The statistics of initial density field

We assume that the initial overdensity field, $\delta(\mathbf{x}) \equiv [\rho(\mathbf{x}) - \bar{\rho}]/\bar{\rho}$ (whose Fourier transform is denoted by $\delta_{\mathbf{k}}$), is Gaussian and is described by a power spectrum $P(k) \equiv \langle |\delta_{\mathbf{k}}|^2 \rangle$. For most of our discussions, we will take the standard CDM model (Blumenthal et al. 1984; Davis et al. 1985) as an example. In this model the power spectrum of the mass density fluctuations is given by equation (G3) of BBKS:

$$P(k) = A\kappa T^2(\kappa); \tag{1}$$

$$T(\kappa) = \frac{\ln(1 + 2.34\kappa)}{2.34\kappa} \left[1 + 3.89\kappa + (16.1\kappa)^2 + (5.46\kappa)^3 + (6.71\kappa)^4\right]^{-1/4}$$

where $\kappa \equiv k/[(\Omega h^2)\mathrm{Mpc}^{-1}]$. We take $\Omega = 1$ and $h = 0.5$. $A$ is a normalization factor to be specified below. We will also consider scale-free models with power-law spectra:

$$P(k) = Ak^n. \tag{2}$$

The field $\delta(\mathbf{x})$ can be smoothed by convolving it with a spherical symmetric window function $W(R_0; r)$ having *comoving* radius $R_0$ (measured in current units). The smoothed field is

$$\delta(R_0; \mathbf{x}) = \int W(R_0; |\mathbf{x} - \mathbf{y}|)\delta(\mathbf{y})d\mathbf{y}$$

$$= \sum_{\mathbf{k}} \delta_{\mathbf{k}} \hat{W}(R_0; k) \exp(i\mathbf{k} \cdot \mathbf{x}), \tag{3}$$

where $\hat{W}(R_0, k)$ is the Fourier transform of the window function $W(R_0; r)$. Therefore, for each point $\mathbf{x}$, there is a trajectory $\delta_0 = \delta(R_0)$ which describes the overdensity $\delta$ as a function of the window radius $R_0$. A useful quantity characterising the power spectrum is the rms fluctuation of mass in a given smoothing window:

$$\Delta^2(R_0) = \langle |\delta(R_0; \mathbf{x})|^2 \rangle = \sum_{\mathbf{k}} P(k)\hat{W}^2(R_0; k). \tag{4}$$

A convenient way to normalize the power spectrum [i.e. to determine $A$ in equations (1) and (2)] is to specify $\Delta(R_0)$ at a given radius for a given window function. We write $\Delta(8\,h^{-1}\mathrm{Mpc}) = \sigma_8$ for a top-hat window: $W(R_0; r) = 1$ if $r \leq R_0$ and 0 otherwise. This normalization is related to the conventional "bias factor" $b$ through $\sigma_8 = 1/b$.

For a given window function the smoothed field $\delta(R_0; \mathbf{x})$ is Gaussian and obeys the following distribution function

$$p(\nu_0)d\nu_0 = \frac{1}{(2\pi)^{1/2}} \exp^{-\nu_0^2/2} d\nu_0, \qquad (5)$$

where $\nu_0 \equiv \delta_0/\Delta(R_0)$. Since both $\delta_0$ and $\Delta$ grow with time in the same manner in linear perturbation theory, it is convenient to use $\delta_0$ and $\Delta$ linearly extrapolated to present time. It is clear that the extrapolated quantities still obey equation (5). In what follows, we write our formula in terms of the extrapolated quantities, unless otherwise stated. Also we will omit writing explicitly the smoothing radius $R_0$, but we will often use subscripts to distinguish $\Delta$, and other quantities, at different smoothing lengths [e.g. $\Delta_0 \equiv \Delta(R_0)$, $\Delta_1 \equiv \Delta(R_1)$]. For a top-hat window function, which is adopted for almost all our discussion, the average mass $M(R_0)$ contained in a window of radius $R_0$ is $\bar{M}_0 = (4\pi/3)\rho_0 R_0^3$, where $\rho_0$ is the mean density of the universe. For a given spectrum, the quantities $R_0$, $\Delta_0$ and $\bar{M}_0$ are equivalent variables.

The two-point correlation function of mass in Lagrangian space, $\xi_{\rm m}^{\rm L}$, can now be readily derived from the statistics of the Gaussian density field. The average mass correlation function $\bar{\xi}_{\rm m}^{\rm L}$, which is related to the two-point correlation $\xi_{\rm m}^{\rm L}$ by

$$\bar{\xi}_{\rm m}^{\rm L}(R_0) \equiv \frac{1}{V_0} \int_{V_0} 4\pi r^2 \, dr \xi_{\rm m}^{\rm L}(r) \qquad (6)$$

with $V_0 = 4\pi R_0^3/3$, can be formally written as

$$\bar{\xi}_{\rm m}^{\rm L}(R_0) = \frac{1}{\bar{M}_0} \int M_0 p(\Delta_0, \delta_0|m) d\delta_0 - 1 \qquad (7a)$$

where $M_0 = (1 + \delta_0)\bar{M}_0$, and $p(\Delta_0, \delta_0|m)$, which we abbreviate as $p(0|m)$ hereafter, is the conditional probability for a spherical region with comoving radius $R_0$ to have a mean linear overdensity $\delta_0$, given that there is a mass particle in its central volume element. According to Bayes' theorem, we write $p(0|m) \propto p(0)p(m|0)$ where $p(0)$ is an abbreviation of $p(\nu_0)$ given by equation (5); $p(m|0)$ is the probability of finding a mass particle in the central volume element of a sphere with radius $R_0$ and overdensity $\delta_0$. We expect that $p(m|0)$ is proportional to the particle number density at the center of the sphere, or approximately $p(m|0) \propto M_0/R_0^3$. $\bar{\xi}_{\rm m}^{\rm L}(R_0)$ can then be written as

$$\bar{\xi}_{\rm m}^{\rm L}(R_0) = \frac{1}{\bar{M}_0} \frac{\int M_0^2 p(0) d\delta_0}{\int M_0 p(0) d\delta_0} - 1. \qquad (7b)$$

In the present case $M_0 = \bar{M}_0(1 + \delta_0)$, and $\bar{\xi}_{\rm m}^{\rm L}(R_0) = \int \delta_0^2 p(0) d\delta_0 = \Delta^2(R_0)$. This result is trivial, but the procedure is quite suggestive: if a quantity (in the above example, the mass) is a function of the linear density $\delta_0$ in a given window, it is possible to calculate the spatial correlation function of this quantity. In the next subsection, we will see that the number of dark haloes in a region is related to the linear overdensity of that region. We can therefore hope to obtain an analytical model for the correlation functions of dark haloes in Lagrangian space.

## 2.2 Dark haloes from the initial density field

We assume that dark haloes are spherical symmetric, virialized clumps of dark matter. In an Einstein-de Sitter universe, the physical radius $R$ of a spherically symmetric perturbation with comoving initial radius $R_0$ and (extrapolated) mean interior density contrast $\delta_0 > 0$ evolves with redshift $z$ as

$$\frac{R(z)}{R_0} = \frac{3}{10} \frac{1 - \cos\theta}{\delta_0}; \tag{8a}$$

$$(1 + z)^{-1} = \frac{3 \times 6^{2/3}}{20} \frac{(\theta - \sin\theta)^{2/3}}{\delta_0}. \tag{8b}$$

So the radius of a perturbation will reach its maximum $R_{\rm m}$ at redshift $z_{\rm m}$, with $R_{\rm m}$ and $z_{\rm m}$ given by equation (8) with $\theta = \pi$. The perturbation collapses to a point ($R = 0$) at a redshift $z_c = 1.686/\delta_0 - 1$. In practice, the perturbation will not, of course, collapse to a point, but will virialize at a radius $R_{\rm vir}$. It is usually assumed that a collapsing structure virializes at half its radius of maximum expansion. This would give a density contrast at the time of collapse of $\sim 178$. We model the density profile of dark haloes by that of a singular isothermal sphere. The mass $M$ and circular velocity $v_c$ [defined by $v_c = (GM/r)^{1/2}$] of a halo are therefore related to its initial comoving radius $R_1$ (note: in what follows the properties of dark haloes are labelled by subscripts 1, 2; the subscript 0 is reserved for the properties of uncollapsed spherical regions) and the redshift $z$ at which it is found by

$$M = \frac{4\pi}{3}\rho_0 R_1^3; \quad v_c = 1.67(1 + z)^{1/2} H R_1, \tag{9}$$

where the latter equation assumes that the mean density contrast of clumps when they virialize is 178 and that the haloes found at redshift $z$ have all just virialized.

The probability for a spherical region to be part of a single collapsed structure (i.e. for its overdensity $\delta$ to exceed $\delta_c \equiv 1.686$) by redshift $z$ is

$$F(R_1, z) = \int_{\delta_c}^{\infty} p(R_1, z, \delta) d\delta, \qquad (10)$$

where $p(R_1, z, \delta)$ is given by equation (5) with $R_0$ and $\delta_0$ replaced by $R_1$ and $\delta(1+z)$ respectively. According to PS, $F(R_1, z)$ gives half the fraction of matter which is in haloes of radius exceeding $R_1$ (or mass exceeding $M$, see equation 9 above) at redshift $z$ (see also Bond et al. 1991, hereafter BCEK). The differential mass distribution is then

$$f(M, z) dM = -2 \frac{\partial F}{\partial M} dM = \frac{2}{(2\pi)^{1/2}} \frac{\delta}{\Delta_1^2} \exp\left[-\frac{\delta^2}{2\Delta_1^2}\right] \frac{d\Delta_1}{dM} dM. \qquad (11)$$

Hence the comoving number density of haloes, expressed in current units, as a function of $v_c$ and $z$ is

$$n(v_c, z) dV_c = \frac{-3(1.67^3)\delta_c H^3(1+z)^{5/2}}{(2\pi)^{3/2} v_c^4 \Delta(R_1)} \frac{d\ln\Delta}{d\ln v_c} \exp\left[-\frac{\delta_c^2(1+z)^2}{2\Delta^2(R_1)}\right] dv_c \qquad (12)$$

(see White and Frenk, 1991 for details). Since equation (12) applies to haloes that have not been incorporated into larger collapsed systems at a given redshift, it accounts automatically for the cloud-in-cloud problem.

We also need a formula to describe the relation between haloes and the surrounding density field. BCEK show that the probability that a randomly chosen spherical region $(R_0, \delta_0)$ which is *not* contained in a collapsed object at redshift $z$ is

$$q(\nu_0) d\nu_0 = \frac{1}{(2\pi)^{1/2}} \left[e^{-\nu_0^2/2} - e^{-(\nu_0 - 2\nu_c)^2/2}\right] d\nu_0, \qquad (13)$$

where $\nu_c \equiv \delta_c(1+z)/\Delta_0$. The fraction of the mass in such regions of mass $M_0$ (corresponding to rms overdensity $\Delta_0$) and present extrapolated overdensity $\delta_0$, which at redshift $z_1$ [corresponding to extrapolated critical overdensity $\delta_1 = (1+z_1)\delta_c$] is in haloes with mass in the range $M_1 \to M_1 + dM_1$ (where $M_1 < M_0$ by definition) is

$$f(\Delta_1, \delta_1 | \Delta_0, \delta_0) \frac{d\Delta_1^2}{dM_1} dM_1 = \frac{1}{(2\pi)^{1/2}} \frac{\delta_1 - \delta_0}{(\Delta_1^2 - \Delta_0^2)^{3/2}} \exp\left[-\frac{(\delta_1 - \delta_0)^2}{2(\Delta_1^2 - \Delta_0^2)}\right] \frac{d\Delta_1^2}{dM_1} dM_1. \qquad (14)$$

(Bower 1991; BCEK). So the average number of $M_1$ haloes at redshift $z_1$ in a spherical region with comoving radius $R_0$ and overdensity $\delta_0$ is

$$\mathcal{N}(1|0) \frac{d\Delta_1^2}{dM_1} dM_1 \equiv \frac{M_0}{M_1} f(1|0) \frac{d\Delta_1^2}{dM_1} dM_1. \qquad (15)$$

where $f(1|0) \equiv f(\Delta_1, \delta_1 | \Delta_0, \delta_0)$. Notice that since $M_1$ is collapsed at $z_1 > 0$ whereas $M_0$ is analysed at $z < z_1$, we have that $\delta_1 > \delta_0$. For haloes of a given mass $M_1$ and redshift $z_1$, the distribution of this number is determined by the statistics of the background density field $(\Delta_0, \delta_0)$.

## 2.3 Correlations of dark haloes in Lagrangian space

Let us first consider the cross-correlation between dark haloes and mass. Denote by $p(\Delta_0, \delta_0 | \Delta_1, \delta_1) d\delta_0$ or simply $p(0|1) d\delta_0$ the conditional probability for a spherical region with comoving radius $R_0$ to have a mean linear overdensity $\delta_0$, given that there is, at redshift $z_1 = \delta_1/\delta_c - 1$, a dark halo with mass $M_1$ (corresponding to initial comoving radius $R_1 < R_0$) at its centre. To be consistent with the definition that a halo should not be contained in a larger halo, the region $(R_0, \delta_0)$ should also not be contained in a larger collapsed region. The average cross-correlation function between haloes and mass, at radius $R_0$ in Lagrangian space, $\bar{\xi}_{\mathrm{hm}}^{\mathrm{L}}(R_0)$, can formally be written as

$$\bar{\xi}_{\mathrm{hm}}^{\mathrm{L}}(R_0) = \int_{-\infty}^{(1+z_1)\delta_c} \delta_0 p(0|1) d\delta_0, \tag{16}$$

where the integral limit is chosen so that the larger region has not collapsed by redshift $z_1$. According to Bayes' theorem, we write

$$p(0|1) \propto q(0) p(1|0) \tag{17}$$

where $q(0)$ is given by equation (13); $p(1|0)$ is the probability to find a halo of type 1 (i.e. with corresponding initial redius $R_1$ and identified at redshift $z_1$) at the centre of a Lagrangian sphere with radius $R_0$ and with overdensity $\delta_0$. We assume that $p(1|0)$ is proportional to the number density of type 1 haloes at the center of the spherical region, and write $p(1|0) \propto \mathcal{N}(1|0)/R_0^3$. This is an approximation, because it assumes that the central number density of haloes is equal to the average number density. The function $\bar{\xi}_{\mathrm{hm}}^{\mathrm{L}}(R_0)$ can then be written as

$$\bar{\xi}_{\mathrm{hm}}^{\mathrm{L}}(R_0) = \frac{\int_{-\infty}^{(1+z_1)\delta_c} \delta_0 q(0) \mathcal{N}(1|0) d\delta_0}{\int_{-\infty}^{(1+z_1)\delta_c} q(0) \mathcal{N}(1|0) d\delta_0}. \tag{18}$$

It can be proven that the denominator in equation (18) is equal to $n(v_1, z_1) V_0$, where $V_0 \equiv 4\pi R_0^3/3$ and $n(v_1, z_1)$ is the comoving density of type 1 haloes (equation 12).

The cross-correlation function between haloes of type 1 and those of type 2 in Lagrangian space, $\bar{\xi}_{12}^{L}(R_0)$, can be obtained in a similar way. Let $\mathcal{N}(2|0;1)$ be the number of type 2 haloes in a spherical region with comoving radius $R_0$, given that this region has an overdensity $\delta_0$ and a halo of type 1 in its central volume element. We can write

$$\bar{\xi}_{12}^{L}(R_0) = \frac{1}{n(v_2, z_2)V_0} \int_{-\infty}^{(1+z_1)\delta_c} \mathcal{N}(2|0;1) p(0|1) d\delta_0 - 1 \tag{19}$$

Since haloes are spatially exclusive in our model, we assume that

$$\mathcal{N}(2|0;1) = \frac{M_0 - M_1}{M_2} f(\Delta_2, \delta_2 | \Delta_0, \delta_0). \tag{20}$$

Using the same argument as in deriving equation (18), we obtain

$$\bar{\xi}_{12}^{L}(R_0) = \frac{1}{n(v_2, z_2)V_0} \frac{\int_{-\infty}^{(1+z_1)\delta_c} \mathcal{N}(2|0;1)\mathcal{N}(1|0)q(0)d\delta_0}{\int_{-\infty}^{(1+z_1)\delta_c} \mathcal{N}(1|0)q(0)d\delta_0} - 1. \tag{21}$$

Since by definition $p(0|1) = 0$ if $\delta_1 < \delta_0$, the integral limits ensure that the large region has not collapsed at $z > \min(z_1, z_2)$. The $\bar{\xi}_{12}^{L}$ so defined has the desired property that $\bar{\xi}_{12}^{L} = \bar{\xi}_{21}^{L}$, if $M_0$ is much larger than both $M_1$ and $M_2$.

### 2.4 Dynamical evolution of correlation functions

The correlation functions defined by equations (7b), (18) and (21) are correlation functions in Lagrangian space. In these cases we have an ensemble of spheres with the same Lagrangian radius, and the correlation functions are estimated by the statistics of masses and halo numbers in these spheres. The correlation functions in physical (Eulerian) space should, however, be estimated from samples of spheres with the same *physical* radius. We therefore need to describe the dynamical evolution of the correlation functions. To treat the dynamical evolution accurately one needs numerical simulations. Here we propose an analytic model based on simple approximations which we discuss below.

Within a spherical region with radius $R_0$ in Lagrangian space, the number of haloes which have mass in the range $M_1 \to M_1 + dM_1$ at redshift $z_1$ is given, via the mean overdensity of the region $\delta_0$, by equation (15). We assume that the mass shell $(R_0, \delta_0)$ will contract or expand, depending on $\delta_0 > 0$ or $< 0$, according to the spherical evolution model of density perturbations in an Einstein-de Sitter universe. We also assume that as the mass shell contracts or expands the mass and the number of dark haloes within it do not change. The conservation of mass is easy to understand, while the conservation of the number of dark haloes is justified by the fact that the definition of dark haloes involved in equations (12)-(15) takes account the disappearance of dark haloes due to merging. The evolution of the radius $R$ of a mass shell $(R_0, \delta_0)$ with $\delta_0 > 0$ is described by equation (8). For $\delta_0 < 0$, we should replace $(1 - \cos\theta)$ in equation (8a) by $(\text{ch}\theta - 1)$, $(\theta - \sin\theta)$ in equation (8b) by $(\text{sh}\theta - \theta)$, and $\delta_0$ by $|\delta_0|$. For a given redshift, a physical radius $R$ corresponds to a curve in the $(R_0, \delta_0)$ space. In Figure 1 we plot three (solid) curves which correspond to $R = 1$, 10 and 100 Mpc at $z = 0$. Since the evolution equations depend on $R$ through $R/R_0$, curves corresponding to other values of $R$ can be obtained by shifting the curves on the plot along the $R_0$ axis. These curves represent the evolution of different mass shells before collapsing (i.e. $\delta_0 < \delta_c$). Each point on the curves corresponds to a spherical mass shell which evolves into the corresponding physical radius at $z = 0$ (or similarly at another redshift). Now suppose we have an ensemble of spherical regions with radius $R$ in *Eulerian* space, each of which corresponds to a point in the $R_0$–$\delta_0$ space. Since the mass and the number of haloes contained in each sphere are known, one can estimate the correlation functions at a physical radius $R$ from the statistics of the mass and halo numbers.

Let $p_E(\delta_0|1, R)d\delta_0$ be the probability of finding a spherical region with Eulerian radius $R$ and with linear overdensity in the range $\delta_0 \to \delta_0 + d\delta_0$, given that there is a type-1 halo at its centre. We can relate the Lagrangian radius $R_0$ of this region to $R$ and $\delta_0$ through the spherical model of perturbation evolution. It is obvious that accuracy of the result will depend on how accurate it is to use the spherical model in describing the evolution of a mass shell $(R_0, \delta_0)$. We may expect that the spherical model should work better for mass shells that have a larger mass halo at their centers. We therefore write the average cross-correlation functions between type 1 haloes and mass and between type-1 and type-2 haloes in Eulerian space as

$$\bar{\xi}^E_{hm}(R) = \int_{-\infty}^{(1+z_1)\delta_c} \left[\left(\frac{R_0}{R}\right)^3 - 1\right] p_E(\delta_0|1, R)d\delta_0; \tag{22}$$

and

$$\bar{\xi}^E_{12}(R) = \frac{1}{n(\nu_2, z_2)V} \int_{-\infty}^{(1+z_1)\delta_c} \mathcal{N}(2|0;1) p_E(\delta_0|1, R)d\delta_0 - 1, \tag{23}$$

respectively, where $V \equiv (4\pi/3)R^3$. In equation (23), we assume, without loss of generality, that $M_1 \geq M_2$. The factor $[(R_0/R)^3 - 1]$ in equation (22) gives the density contrast in Eulerian space.

It is important to note that the probability $p_{\rm E}(\delta_0|1, R)$ is defined for a halo at the centre of an Eulerian sphere. According to Bayes' theorem, we can write

$$p_{\rm E}(\delta_0|1, R) = \frac{p_{\rm E}(1|0)p(\delta_0|R)}{\int p_{\rm E}(1|0)p(\delta_0|R)d\delta_0}, \qquad (24a)$$

where $p(\delta_0|R)d\delta_0$ is the probability that a spherical region with Eulerian radius $R$ has linear overdensity in the range $\delta_0 \to \delta_0 + d\delta_0$, and $p_{\rm E}(1|0)$ is the probability of finding a halo of type 1 at the centre of such a sphere. As before, we assume that $p_{\rm E}(1|0)$ is proportional to the number density of type-1 haloes, but now in Eulerian space. Then we have

$$p_{\rm E}(1|0) \propto \frac{1}{R^3} \mathcal{N}(1|0). \qquad (24b)$$

Equation (24b) is, of course, only an approximation, because it assumes that the number density of haloes at the center of the spherical region is proportional to the average number density within the region. We can now write

$$p_{\rm E}(\delta_0|1, R) = \frac{\mathcal{N}(1|0)p(\delta_0|R)}{\int_{-\infty}^{(1+z_1)\delta_c} \mathcal{N}(1|0)p(\delta_0|R)d\delta_0}. \qquad (25)$$

To derive an expression for $p(\delta_0|R)$, we consider the corresponding cumulative function $p(\delta_0 > \delta|R)$. For the power spectra under consideration, the rms mass fluctuation $\Delta(R_0)$ decreases monotonically with increasing $R_0$. The quantity $\nu_0 \equiv \delta_0/\Delta_0$, therefore, increases monotonically with $\delta_0$ for a given $R$. Our problem now becomes to find $p(\nu_0 > \nu|R)$. We note that, for a given $R$, $\nu_0$ goes from $-\infty$ to $\infty$ as $\delta_0$ goes from $-\infty$ to $\delta_c \equiv 1.686$. For a Gaussian field, $\nu_0$ obeys equation (5) for a given Lagrangian radius. We therefore make the following *ansatz*:

$$p(\nu_0 > \nu|R) = \int_\nu^\infty \frac{1}{\sqrt{2\pi}} e^{-\nu_0^2/2} d\nu_0. \qquad (26a)$$

This ansatz implies that, for given $R$ and $z$, the function $p(\delta_0|R)$ is given by

$$p(\delta_0|R) = \frac{1}{\sqrt{2\pi}} e^{-\nu_0^2/2} \frac{d\nu_0}{d\delta_0}, \qquad (26b)$$

where $\nu_0 = \delta_0/\Delta(R_0)$ depends on $\delta_0$ both directly and through $R_0 = R_0(\delta_0, R)$ which is given by the spherical model of equation (8).

The motivation for our ansatz is as follows. We recall that, for each point $\mathbf{x}$ in the Lagrangian space, the overdensity $\delta_0$ in a window of Lagrangian radius $R_0$ is described by the trajectory $\delta_0(R_0)$ of a "particle" in $R_0$–$\delta_0$ space (see Figure 1). All trajectories obey the boundary condition that $\delta_0 \to 0$ when $R_0 \to \infty$. As shown in Figure 1, for a given Eulerian radius $R$, each trajectory will pass through $R_0 = R$ as the "particle" moves from $R_0 = \infty$ to $R_0 = 0$. Each trajectory will also cross the boundary $\delta_0 = \delta_0(R_0, R)$, since the probability for a particle to go to $\delta_0 = -\infty$ is zero. For a given $R_0$, the probability for a trajectory to cross the vertical line $R_0 = R_0$ above $\delta_0 = \delta$ is given exactly by equation (26a) with $\nu = \delta/\Delta_0$. If each trajectory crossed the boundary $\delta_0 = \delta_0(R_0, R)$ only once, this would also be the probability for a trajectory to cross the boundary above $\delta_0 = \delta$, and the ansatz (26a) would be exact. However, a trajectory can cross a given boundary more than once in two different ways: it can either cross the boundary more than once at $\delta_0 < \delta_c$, or cross the boundary first at $\delta_0 = \delta_c$ and then at $\delta_0 < \delta_c$. These two cases are schematically shown in Figure 1 by short-dashed and dotted curves, respectively, for a point 'A' on the boundary $\delta_0 = \delta_0(R_0, R)$ with $R = 10$ Mpc. These kinds of trajectories should be excluded in calculating the correlation functions of dark haloes. In the first case, the contribution of point 'A' to the correlation function at Eulerian radius $R = 10$ Mpc has already been taken into account at point 'C', because the trajectory implies that the region represented by 'A' is contained in that represented by 'C'. In the second case, the region 'A' is part of a larger region (represented by 'B') which has collapsed at $z = 0$ (because the overdensity within it has reached $\delta_0 = \delta_c$) and no haloes can exist in such a region according to our definition of dark haloes. According to BCEK, the probability of finding a spherical region $(R_0, \delta_0)$ which is not a part of a larger collapsed region is given by equation (13). This suggests that it may be better to make our ansatz for $p(\delta_0|R)$ according to equation (13) instead of equation (5). As we will show later, both ansätze give quite similar results, and the ansatz based on equation (13) is not superior to that based on (5) for a reason we discuss below.

To see how well our ansatz works, we generate a large number of trajectories in $(R_0, \delta_0)$ space (see Figure 1) starting from $R_0 = 1000$ Mpc (the results do not change significantly if we start from a larger $R_0$) and $\delta_0 = 0$. We calculate the fraction of such trajectories which first cross the boundary $\delta_0 = \delta_0(R_0, R)$ for a given $R$ near $\delta_0 = \delta$. This fraction gives the probability $p(\delta_0|R)$. The trajectories are generated using the fact that, for a sharp $k$-space window function $[\hat{W}(R_0, k) = 1$ if $k < 1/R_0$ and $\hat{W}(R_0, k) = 0$ otherwise], the change in $\delta_0$, $D\delta_0$, due to a change of $R_0$ from $R_0$ to $R_1 \equiv R_0 - DR_0$, is a Markov random walk governed by a probability function

$$p(D\delta_0) = \frac{1}{[2\pi(\Delta_1^2 - \Delta_0^2)]^{1/2}} \exp\left[-\frac{(D\delta_0)^2}{2(\Delta_1^2 - \Delta_0^2)}\right]. \qquad (27)$$

(see BCEK for a discussion). For a top-hat filter as used in the present paper, equation (27) is only an approximation since successive steps in the random walk are correlated (Bower 1991). In Figure 2 we compare the results given by our ansatz (26) (solid curves) and that based on equation (13) (dashed curves) to those given by the Monte Carlo trajectories described above. The results shown are for a $\sigma_8 = 1$ CDM spectrum and for two values of Eulerian radius, $R = 10$ and 2 Mpc, representing large and small scales. The figure shows that both models are quite accurate for $R \gtrsim 10$ Mpc and that they are indistiguishable on these scales (note: the larger $R$ is, the more accurate the models are). The ansatz (26) holds also reasonably well on scales as small as 2 Mpc, but it tends to give excessive weight to regions with low linear densities. This is clearly due to fact that for a given $R$ a lower value of $\delta_0$ corresponds to a lower value of $R_0$ and a larger value of $\Delta_0$. A trajectory thus has a larger probability of crossing a given boundary $\delta_0 = \delta_0(R_0, R)$ several times for smaller values of $R_0$. As a result the ansatz overestimates the probability of *first* crossing at lower value of $\delta_0$. This overestimation is more severe in the model based on equation (13), because this model does not allow multiple crossings, and so reduces the relative probability of high values of $\delta_0$. The accuracy of the ansatz may depend on the amplitude and shape of the power spectrum. For a power spectrum in which the rms fluctuation $\Delta$ on small scales is larger than that for the $\sigma_8 = 1$ CDM spectrum, the ansatz works worse for small $R$. To get an idea of how strongly our results depend on the ansatz, we show in Figure 3 the correlation functions obtained by using model (26) (curves) and that based on equation (13) (crosses). The results are shown for haloes with different circular velocities (or masses, see equation 9) in a CDM model with $\sigma_8 = 1$. The two models agree very well for most cases. Noticeable disagreement appears only on small scales for haloes with low circular velocities ($v_c < 100 \, \mathrm{km \, s^{-1}}$). The agreement is better for CDM models with smaller $\sigma_8$, because this means a lower value of $\Delta(R_0)$ for a given $R_0$. We therefore believe that the difference between the ansatz (26) and the proper probability function given by the Monte-Carlo simulation is unlikely to produce a substantial difference in the results; we use this ansatz in the remainder of our discussion.

A similar model can be constructed for the autocorrelation of the mass by using equation (7b). Since the mass correlation depends also on the virialization of small structures, we will present this model in a separate paper.

## 3 COMPARISON WITH NUMERICAL SIMULATIONS

The critical test of our analytic model comes from a comparison with numerical sim-ulations. Here we compare our model predictions for the two-point correlation functions of dark haloes with the results given by numerical simulations.

Gelb and Bertschinger (1994a, hereafter GB94a) have carried out high resolution N-body simulations for CDM models with $\Omega = 1$, $h = 0.5$ and different bias parameters. GB94a used two algorithms to identify haloes. The first algorithm (called DENMAX) identifies haloes as local density maxima in the smoothed, evolved density field. In the second algorithm (the friends-of-friends algorithm, called FOF), haloes are selected from the evolved particle positions by identifying all particles within a given linking distance ($l$) of each other. Two linking distances were used, with $l = 0.1$ and $0.2$ times the mean particle separation in the simulation. As discussed in detail in GB94a, the DENMAX may have an advantage over FOF in its ability to break up large dense clusters into subgroups, while still being able to detect smaller, less dense haloes in the field. In Figure 4 the solid, short-dashed and long-dashed curves show the average correlation functions of haloes identified by DENMAX, FOF($l = 0.1$) and FOF($l = 0.2$), respectively. Results are shown for CDM models with $\sigma_8 = 1$ (4a) and $\sigma_8 = 0.5$ (4b). Each curve shows the correlation function of haloes with masses (or circular velocities) larger than some threshold value. This value, which depends on the algorithm for halo identification, is chosen so that the number density of haloes selected from the simulation is the same as that predicted by equation (12) for haloes with $v_c$ greater than a given threshold value $V_c$ (indicated in each panel). The number of haloes in the simulation box, which has volume $(100 \mathrm{Mpc})^3$, is indicated by $N_\mathrm{h}$. The predictions of our analytic model for the same model parameters are shown as crosses. The average correlation function between these haloes are obtained from equation (23) by integrating both the denominator and the numerator in the first term on the right hand side over $v_1$ and $v_2$ (which are related to $M_1$ and $M_2$ by equation 9) from $V_c$ to $\infty$. The dotted lines in the figure show the average correlation function which corresponds to the differential correlation function $\xi(r) = [5\,h^{-1}\mathrm{Mpc}/r]^{-1.8}$ observed for normal galaxies. The first thing one notices in this figure is the dependence of the correlation function on the halo identification algorithm; the simulation with DENMAX gives the strongest correlations and FOF($l = 0.2$) the weakest. The effect is quite large on small scales, and it is stronger for bigger haloes. Such a dependence is obviously due to the fact that the FOF algorithm with large linking length tends to merge haloes in high density regions and thereby reduce the weight of such regions (see GB94a for a discussion). The other remarkable result shown in the figure is the agreement between the correlation function predicted by our model and that based on the DENMAX algorithm. For $\sigma_8 = 1$, the agreement is almost perfect. For $\sigma_8 = 0.5$, our model gives a higher amplitude on small scales for haloes with $v_c \gtrsim 200\,\mathrm{km\,s^{-1}}$. Given the many assumptions made by our model and the uncertainties inherent to the simulation, we regard this agreement to be as good as we could hope for. The reasons for this agreement are far from clear, because, as we will show in Section 4, the low-mass haloes do not correspond to high peaks of the initial density field and our correlation functions differ very substantially from their Lagrangian analogues.

It is interesting to compare our results with those given by Katz, Quinn and Gelb (1993). They used the same set of simulations to compare the correlation function of particles initially located near peaks in the linear density field with that of the haloes which actually form. They found the correlation function of the tagged peaks to be much larger than that of the haloes. Similar results were found by Mann, Heavens & Peacock (1993), based on an analytic model for the clustering of peaks. That our model does a better job suggests that the PS formalism indeed solves, at least partially, the cloud-in-cloud problem. Of course, it is unclear that this is any advantage when a comparison is to be made with the observed *galaxy* correlation function, since the haloes which represent galaxy clusters should each contain several galaxies, an effect which may be better represented by the peak model.

Figure 5 shows the results for scale-free models with $n = -1.5$ (5a) and 0 (5b). The N-body simulations used here are similar to those in EFWD, but with a larger number of particles ($10^6$) and a higher force resolution ($L/2500$ where $L$ is the side of the computational box). The values of the scale factor $a$ given on the figure give the expansion of the simulation since its initial condition. The initial power spectrum was normalized as described in EFWD. We compare analytic model with simulation in the same manner as in Figure 4. The long-dashed and short-dashed curves show the results for haloes selected by the FOF algorithm with $l = 0.2$ and $0.1$, respectively. The solid curves show the results for haloes selected by a third algorithm which mimics the definition of haloes in the PS formalism. In this algorithm, the centres of haloes are determined by the FOF algorithm with a small linking length $l = 0.1$. The mass of each halo is taken to be the total number of particles in a bounding sphere within which the mass overdensity is 200. The dependence of the correlation function on the algorithm of halo identification is similar to that shown in Figure 4. For the scale-free model with $n = -1.5$, our model prediction (crosses) agrees very well with the results based on the third algorithm. For the $n = 0$ model, our model may fail to fit the results for large haloes at separations where $\bar{\xi} \gtrsim 10$. At larger separations the agreement is again reasonably good.

To test our formalism for haloes with larger mass, we use the simulation results of Bahcall and Cen (1992, BC). BC have carried out large simulations to study the correlation functions of clusters of galaxies. Haloes are selected by an adaptive FOF algorithm which uses smaller linking lengths for particles in higher density regions. The relation of this algorithm to either the original FOF or DENMAX is not obvious. It is, however, plausible that this algorithm should be more effective than FOF in resolving haloes in high density regions. Motivated by observation, BC considered the correlation length $r_0$, defined to be the scale where the two-point correlation function is unity, as a function of the mean separation of clusters $d$ (defined by the mean number density of clusters $n$ by $d \equiv n^{-1/3}$). Their result for a CDM model with $\sigma_8 = 0.75$ is shown in Figure 6 as the dashed curve. For comparison, we also show in the figure the observational data, adopted again from BC. To compare our model with their simulations, we calculate the mean number of haloes with circular velocities $v_c > V_c$, $n(v_c > V_c)$, by equation (12), and estimate the mean separation of the haloes by $d = n^{-1/3}$. To obtain $r_0$, we assume that the average correlation functions have a power-law form $\bar\xi \propto r^{-2}$ around $\bar\xi = 1$. The prediction of our formula for the same CDM model is shown in Figure 6 as the solid curve. The agreement with the numerical simulations is gratifying.

The strong dependence of the above correlation function on the halo identification algorithm may explain why different investigators sometimes get different correlation functions from similar simulations.

## 4 BIAS IN THE DISTRIBUTION OF DARK HALOES

Having shown that our analytic model gives a reasonable approximation to the correlation function of the dark haloes identified in N-body simulations, we now use this model to investigate how the spatial distributions of dark haloes with different circular velocities are related to the mass distribution. This should help us to understand how different kinds of object may be used to trace the mass distribution in the real universe.

### 4.1 Cross-correlation between haloes and mass

Figure 7 shows the Lagrangian space average cross-correlation functions between mass and haloes of differing circular velocity (equation 18). Since the initial density field has been linearly extrapolated to $z = 0$, these cross-correlation functions can be smaller than

−1. Results are shown for CDM models with $\sigma_8 = 1$ and $\sigma_8 = 0.5$, and for dark haloes with circular velocities $v_c = 50, 100, 200, 400, 800, 1600 \,\text{km s}^{-1}$. For comparison, the average mass correlation functions are shown as solid curves. The figure shows that haloes with $v_c \lesssim 800 \,\text{km s}^{-1}$ in the $\sigma_8 = 1$ model, and with $v_c \lesssim 400 \,\text{km s}^{-1}$ in the $\sigma_8 = 0.5$ model, are initially anti-correlated with overdensity. Most of these haloes are located in regions where the initial overdensity is negative. This anti-correlation with mass is stronger for haloes with smaller circular velocity. It occurs because the mass in high density regions has already been incorporated into larger mass haloes by $z = 0$. The effect is also stronger for the model with $\sigma_8 = 1$, because the higher fluctuation amplitude implies substantially larger mass for "typical" haloes. The easist way to understand these results is to use an argument based on a peak-background split (see e.g. Efstathiou et al. 1988). Consider a (large) background region with (extrapolated) overdensity $\delta_0 \ll 1 + z$. In this region the threshold for collapse of a small-scale density peak, $\delta_t \equiv \delta_c(1+z)$, is effectively reduced to $\delta_t - \delta_0$. The comoving number density of halos (equation 12) in this region will be $n(V_c, z, \delta')$ calculated from equation (12) with $\delta_c$ replaced by $\delta'_c \equiv \delta_c - [\delta_0/(1+z)]$. The average background overdensity per halo is

$$\bar{\delta}_0(R_0) \approx \frac{\int_{-\infty}^{\delta_t} \delta_0 n(V_c, z, \delta') p(0) d\delta_0}{\int_{-\infty}^{\delta_t} n(V_c, z, \delta') p(0) d\delta_0}. \tag{28}$$

It is easy to prove that, if $R_0$ encloses a mass much larger than that of the haloes, then equation (18) reduces to equation (30) if we define $\bar{\xi}^{\text{L}}_{\text{hm}}(R_0) = \bar{\delta}_0(R_0)$. If $\delta_0 \ll \delta_t \equiv \delta_c(1+z)$ we have to first order in $\delta_0$

$$n(v_c, z, \delta'_c) = n(v_c, z) \left\{ 1 - \frac{\delta_0}{\delta_t} \left[ 1 - \frac{\delta_t^2}{\Delta^2} \right] \right\}. \tag{29}$$

Substituting in equation (30) shows that haloes with $\delta_t/\Delta > 1$ ($< 1$) will be biased towards regions with $\delta_0 > 0$ ($< 0$) in the initial density field. Unbiased ($v_c^*$) haloes are those for which $\Delta = \delta_t$. For the CDM spectrum (1), $\Delta(R) = \delta_c$ at $R \approx 9$ Mpc for $\sigma_8 = 1$, and at 3 Mpc for $\sigma_8 = 0.5$. Thus, according to equation (9), haloes identified at $z = 0$ and with circular velocities $v_c \lesssim 700 \,\text{km s}^{-1}$ ($\sigma_8 = 1$), $250 \,\text{km s}^{-1}$ ($\sigma_8 = 0.5$) are biased toward regions with $\delta_0 < 0$ in the initial density field. This is exactly what we have seen in Figure 7.

Figure 8 shows the Eulerian cross-correlation functions between haloes and mass given by equation (22) with $z = 0$. Results are shown for the same cases as in Figure 7. However, these Eulerian functions are all positive because of the dynamical evolution of the mass density field. Furthermore, the correlation functions for different mass haloes have a similar shape. The relative bias in these evolved correlation functions is clearly seen, in that haloes with higher $v_c$ are more strongly clustered. The effect is stronger in the model with a lower fluctuation amplitude $\sigma_8$.

### 4.2 Autocorrelation of dark haloes

Figure 9 shows the autocorrelation functions of haloes in Lagrangian space at $z = 0$ obtained from equation (21). Results are shown for CDM models with $\sigma_8 = 1$ and $\sigma_8 = 0.5$, and for haloes of different circular velocity. Unlike the cross-correlation with mass, the amplitudes of the autocorrelation functions in Lagrangian space do not form a monotonic sequence according to increasing $v_c$. The amplitude decreases with increasing $v_c$ until $v_c \sim 800 \, \mathrm{km\, s^{-1}}$ for $\sigma_8 = 1$ ($v_c \sim 200 \, \mathrm{km\, s^{-1}}$ for $\sigma_8 = 0.5$) and then increases with $v_c$. The lowest amplitude is found for haloes whose linear radius $R_1$ (which is related to $v_c$ by equation 9) is such that $\Delta(R_1) \sim \delta_c$, i.e. for present $v_c^*$ haloes. As we have seen in last section, haloes with $\Delta(R_1) > \delta_c$ are anticorrelated with mass initially. The high amplitude of the Lagrangian correlation function for low-mass haloes implies that these haloes are concentrated in lower density regions. The high amplitude for haloes with high $v_c$ means, however, that they are concentrated in high density regions. For $v_c^*$ haloes, the initial clustering is very weak. To see this more clearly, we note that, under the condition that $R_0 \gg R_1$ and $\delta_0 \ll 1$, the correlation function $\bar{\xi}^{\mathrm{L}}(R_0)$ defined by equation (21) with $R_1 = R_2$ and $z_1 = z_2 = z$ can be written as

$$\bar{\xi}_{11}^{\mathrm{L}}(R_0) \approx \frac{\Delta^2(R_0)}{\delta_t^2} \left[ \frac{\delta_t^2}{\Delta_1^2} - 1 \right]^2. \tag{30}$$

For haloes with $\nu_1 \equiv \delta_t/\Delta_1 \gg 1$, $\bar{\xi}_{11}^{\mathrm{L}}(R_0) \approx (\nu_1/\Delta_1)^2 \Delta^2(R_0)$, which shows that the Lagrangian correlation function of haloes with $\nu_1 \gg \Delta_1$ is enhanced with respect to that of mass, $\Delta^2(R_0)$, by a factor of $(\nu_1/\Delta_1)^2$. This result is the same as that for high peaks in the initial density field (see BBKS). For haloes with $\nu_1 \ll 1$, equation (30) gives $\bar{\xi}_{11}^{\mathrm{L}}(R_0) \approx \Delta^2(R_0)/\delta_t^2$, which shows that the Lagrangian correlation function of these haloes is lower than that of mass by a constant factor. For $v_c^*$ haloes with $\nu_1 \sim 1$, $\bar{\xi}_{11}^{\mathrm{L}} \sim 0$. These results are exactly what were shown in Figure 9. The peak-background split may also help to understand the above results (see e.g. Cole 1989). Due to the background modulation, the number of halos at redshift $z$ will be changed by a factor $n(V_c, z, \delta_c')/n(V_c, z, \delta_c)$. The ratio between the perturbation in comoving number density of haloes ($\Delta n/n$) and the (extrapolated) density perturbation $\delta_0$ is

$$\mathcal{R} = -\left[\frac{\partial \ln n(V_c, z, \delta)}{\partial \delta}\right]_{\delta=\delta_c} = \frac{1}{\delta_t}\left[\frac{\delta_t^2}{\Delta_1^2} - 1\right], \qquad (31)$$

which is just the "bias" factor $\sqrt{\xi_{11}^L(R_0)/\Delta^2(R_0)}$.

Figure 10 shows correlation functions for haloes in Eulerian space (equation 23). The results are shown for the same cases as in Figure 9, with each case depicted in the two figures by curves with the same line type. The monotonic increase of the correlation amplitude with $v_c$ is now clearly seen. The dependence is stronger in the model with $\sigma_8 = 0.5$. The shapes of the correlation functions are similar, although those for haloes with larger $v_c$ appear to be steeper. Comparing the results shown in both Figure 9 and Figure 10, we see clearly that the similarity in the shapes of the correlation functions is a result of dynamical evolution rather than of initial conditions. The dynamical evolution of the correlation functions is important for all cases except for haloes with $v_c = 1600\,\mathrm{km\,s^{-1}}$ in the model with $\sigma_8 = 0.5$, where the correlation function in Eulerian space is almost identical to that in Lagrangian space. Thus in "standard CDM" the correlations of rich clusters are almost unaffected by dynamical evolution. Figure 10 also shows that haloes with $v_c > v_c^*$ are more strongly correlated, and those with $v_c < v_c^*$ are less strongly correlated, than mass. This kind of bias in the correlation functions of dark haloes with respect to mass and to each other is also due to dynamical evolutions and differs from the result for high peaks. For example, the amplitude of the correlation function for haloes with $v_c = 400\,\mathrm{km\,s^{-1}}$ is higher than that for haloes with $v_c = 100\,\mathrm{km\,s^{-1}}$ by factors of 2 for $\sigma_8 = 1$, and 4 for $\sigma_8 = 0.5$, while the factor would be about 30 if we use the amplification factor $\nu^2/\Delta^2$ (with $\nu \equiv \delta_c/\Delta$, see equation 30) predicted for high ($\nu \gg 1$) peaks. Thus the "dynamical" bias is weaker than that for high peaks. Also, the relative bias, i.e. the relative enhancement, of the correlation functions for haloes with different $v_c$ depends on the fluctuation amplitude $\sigma_8$, with a lower $\sigma_8$ giving a larger relative bias. This differs from the high peak case, for which the relative bias is constant.

So far we have only considered the correlation function as a function of circular velocity for haloes identified at redshift $z = 0$. It is interesting to see how the present-day correlation function of haloes changes when haloes are identified at higher redshifts. Haloes which were identified at higher redshift may have increased their mass by the present time by accreting, or by merging with other haloes. If galaxies formed in the high redshift haloes and these galaxies kept their identity through the subsequent evolution, the correlation function we now calculate may be a good model for that of galaxies. In Figure 11 we show present-day correlation functions for haloes with $v_c = 200\,\mathrm{km\,s^{-1}}$ which were identified at different redshifts $z$. Results are shown for CDM models with $\sigma_8 = 1$ and with $\sigma_8 = 0.5$. The correlation functions are calculated from equation (23) for $z_1 = z_2 = z$, but with the density perturbation $(\delta_0, R_0)$ evolved to $z = 0$. It is clearly seen that the amplitude of the correlation function increases with $z$, and the effect is stronger for $\sigma_8 = 0.5$. This is a result of the shape of the CDM spectrum on scales $R_1 \sim 1$ Mpc. The rms mass fluctuation on these scales can approximately be written as $\Delta^2(R_1) \propto R_1^{-\gamma}$ with $\gamma \sim 1$ (which corresponds to a power index $n \sim -2$). Since for a given $v_c$, $R_1 \propto (1+z)^{-1/2}$ (see equation 9), we have $\Delta^2 \propto (1+z)^{1/2}$ and $\delta_t^2/\Delta^2 \propto (1+z)^{3/2}$. Using equation (29) we see that haloes identified at higher redshifts are biased toward higher density regions. The trend can be reversed if $n > 1$.

The results presented here show clearly that the present correlation function of dark haloes depends not only on the mass of haloes, but also on the time when they were identified. Indeed, as shown by equation (9), haloes with the same $v_c$ but identified at higher redshift have smaller mass. The results shown in Figure 11 therefore imply that the present-day remnants of early low-mass haloes can be more strongly clustered than present-day haloes of larger mass. This result is interesting. In a hierarchical clustering scenario, such as the CDM models considered here, haloes with high mass formed through the merger of low-mass haloes. If the low-mass haloes had formed galaxies with lower mass, and if these galaxies had neither merged with other galaxies nor significantly increased their masses by accreting the material around them, then these galaxies would be observed today as low mass galaxies with strong clustering. A strong mass (or luminosity) segregation in the correlation function of galaxies, in the sense that galaxies with higher mass (or luminosity) have a stronger correlation function, is not a necessary result of the hierarchical model of structure formation. It is also interesting to note that, for both cases, the correlation of haloes at $z \approx 2$ with a circular velocity $v_c = 200\,\mathrm{km\,s^{-1}}$ may be as strong as that of present-day normal galaxies. A strong 'natural' bias (White et al. 1987) can be present.

### 4.3 Bias

As seen in subsection 4.1, low-mass haloes have initial cross-correlations with mass which are negative. Their positive correlations at $z = 0$ may be due to the motion of both haloes and mass into overdense regions. Thus, a region with $\delta_0 < 0$ and $R_0 > R$ can be in an overdense region with a Eulerian radius $R$, if it is contained in a larger region with $\delta_0 > 0$ and $R_0 = R_0(R, \delta_0)$, while a region with $\delta_0 > 0$ and $R_0 \geq R$ can never be in a underdense region with Eulerian radius $R$. As a result, most of the mass can have moved to overdense regions, although half of it was in underdense regions initially. The same effect could apply to low-mass dark haloes, so that they do not reside in voids at present time, even though they did initially. On the other hand, the positive correlation of low-mass haloes with mass could due primarily to strong evolution of the mass density field around those haloes which reside in high density regions initially, rather than to the evacuation of underdense regions; in this case low-mass haloes would still be biased toward underdense regions. To distinguish between these possibilities, we define a function

$$\delta_{\rm h} = \frac{\mathcal{N}(1|0)}{n(v_1, z_1)V} - 1, \tag{32}$$

for type-1 haloes and for a spherical region with Eulerian radius $R$ (corresponding to a volume $V$) and with present-day mass overdensity $\delta_{\rm m} \equiv (R_0/R)^3 - 1$. This function gives present-day overdensity of type-1 haloes in such a region. In Figure 12a, we show $\delta_{\rm h}$ as a function of $\delta_{\rm m}$ for $R = 20$ Mpc and for haloes (identified at $z = 0$) with different $v_c$. It is clearly seen that present-day haloes with $v_c < v_c^*$ are still biased toward underdense regions.

Figure 12b shows the same thing for $v_c^*$ haloes at varying identification redshift $z$. We see from the figure that at present day all these haloes have correlations comparable to those of the mass. These means that if each $v_c^*$ halo formed a galaxy at high redshift and this galaxy kept its identity through the subsequent evolution, then the distribution of galaxies would not be biased with respect to the mass. A positive bias can be obtained if the majority of normal galaxies form in haloes with $v_c > v_c^*$. In the "standard CDM" model with $\sigma_8 = 0.5$, the circular velocity of $v_c^*$ haloes at $z \gtrsim 1$ is $v_c \lesssim 100\,{\rm km\,s^{-1}}$. Thus, if normal galaxies form only in dark haloes identified at $z \geq 1$ but with circular velocity $v_c \gtrsim 200\,{\rm km\,s^{-1}}$ (typically that of normal galaxies), then the distribution of normal galaxies should be positively biased with respect to the mass, as shown in Figure 11b. The above results suggest that a "natural" bias can be present as a result of constraints on the formation epoch of the dark haloes of normal galaxies. Present-day haloes with $v_c \sim 200\,{\rm km\,s^{-1}}$ may form a different population of galaxies, for example low-surface-brightness galaxies. These galaxies should then have much weaker (by a factor of 2 to 3) correlations than normal galaxies (Mo, McGaugh & Bothun 1994).

## 5 CONCLUSION

In this paper we have developed an analytic formalism for calculating autocorrelation and cross-correlation functions both for dark haloes and for mass. We have tested the results against various N-body simulations. The correlation functions of haloes in the N-body simulations depend strongly on the algorithm by which the haloes are selected. Our formalism agrees reasonably well with results based on algorithms which have stronger ability to break up large clusters than the standard FOF algorithm with a large linking length. We have demonstrated that our formalism can help us to understand how the distributions of different kinds of object are related to that of the mass. Although our discussions are mainly based on CDM models with $\Omega = 1$ and $\Lambda = 0$, the formalism we have developed can easily be extended to other (gaussian) models of structure formation. Also the model can be extended to situations where dark haloes are most appropriately defined in a different way. For example, a similar model can be developed for local maxima in the initial density field. One could also hope to develop similar models for higher order correlation functions, or to develop more realistic models in which the collapse of density perturbations is aspherical.

As pointed out above, it is not straightforward to apply our model to the clustering of galaxies, because a dark halo may contain more than one galaxy, or may not contain galaxies at all. However, if more detailed modelling allows a prediction of the number of galaxies in a halo as a function of the properties of the galaxies and of the halo (see for example, Kauffmann, White & Guiderdoni 1993; Kauffmann, Guiderdoni & White 1994; Cole et al. 1994), the model developed here can readily be extended to study the correlations of galaxies with respect to luminosity, morphological type, colour, or any other property of interest.

## APPENDIX

For readers' convenience, we summarize in the following the main formulae that are needed to calculate the various correlation functions we have discussed. We point out again that these formulae are specific to an Einstein-de Sitter universe. A number of relatively minor changes have to be made in order to apply them to other cosmological models.

For a given spectrum of initial density fluctuation, the rms mass fluctuation in a window with comoving radius $R_0$ (in current units), $\Delta(R_0)$, is given by equation (4). The mass $M$ and circular velocity $v_c$ of a halo identified at redshift $z$ are related to its linear scale $R_1$ by equation (9). The comoving halo number density $n(v_c, z)$, expressed in current units, as a function of $v_c$ (or $R_1$) and $z$ is then given by equation (12).

*The cross-correlation between dark haloes and mass in Lagrangian space* is given by equation (18), with $q(0)$ defined by equation (13) and $\mathcal{N}(1|0)$ by equation (15).

*The cross-correlation between two types of haloes in Lagrangian space* is given by equation (21), with $\mathcal{N}(1|0;1)$ defined by equation (20).

*The cross-correlation between dark haloes and mass in Eulerian space* is given by equation (22), with $p_E$ defined by equation (25). The function $p(\delta_0|R)$ in equation (25) is defined by equation (26b). The evolution of the correlation function is treated by a spherical model which relates the Lagrangian radius $R_0$ of a spherical region to its Eulerian radius $R$ and linear overdensity $\delta_0$ (equation 8).

*The cross-correlation between two types of haloes in Eulerian space* is given by equation (23). If the initial density perturbations (described by mass shells $(\delta_0, R_0)$) are evolved to present time (or any other time), this formula gives the present cross-correlation between the locations of haloes identified at $z_1$ with those identified at $z_2$.

## Acknowledgements

We thank Ed Bertschinger for providing the data on dark haloes in his simulations with Jim Gelb. We also thank Guinevere Kauffmann for help in the selection of haloes in the simulations of scale-free models.

# Figure captions

**Figure 1.** The three solid curves show the evolution of mass shells (defineded by their Lagrangian radius $R_0$ or their mass $M$ and by the mean mass overdensity $\delta_0$ within them) before they collapse, i.e. for $\delta_0 < \delta_c$, where $\delta_c = 1.68$ is shown as the upper horizontal dotted line. These mass shells evolve to have Eulerian radii $R = 1$, 10, or 100 Mpc. These values are shown as the dotted vertical lines. The other three curves show three different random walks representing possible overdensity trajectories that end up predicting halo at point "H" in the $R_0$ - $\delta_0$ space (see the text for details).

**Figure 2.** The probability functions $p(\delta_0|R)$ given by our ansatz (26) (solid curves) and by a variant of it based on equation (13) (dashed curves) are compared to those obtained from Monte Carlo trajectories as described in the text (dotted curves). The results shown are for a CDM spectrum and for two values of the Eulerian radius, $R = 10$ and 2 Mpc.

**Figure 3.** Eulerian average autocorrelation functions for haloes calculated from equation (23) with $p(\delta_0|R)$ given by model (26) (curves) and by a model based on equation (13) (crosses). The results are shown for a CDM model with $\sigma_8 = 1$, and for haloes with different circular velocities $v_c$.

**Figure 4.** Eulerian average autocorrelation functions for haloes with different circular velocities in CDM models with $\sigma_8 = 1$ (4a) and $\sigma_8 = 0.5$ (4b) taken from the numerical simulations of Gelb and Bertschinger (1993a). The solid, short-dashed and long-dashed curves show the results for haloes selected by DENMAX, FOF($l = 0.1$) and FOF($l = 0.2$) respectively. The predictions of our analytic model (equation 23) for the same model parameters are shown as crosses. The dotted lines in the figure show the average correlation function which corresponds to the differential correlation function $\xi(r) = [5\,h^{-1}\mathrm{Mpc}/r]^{-1.8}$ observed for normal galaxies. The value of $N_\mathrm{h}$ in each panel gives the number of haloes in the simulation box.

**Figure 5.** Eulerian average autocorrelation functions for haloes in scale-free models with $n = -1.5$ (5a) and $n = 0$ (5b). The short-dashed and long-dashed curves show the results for haloes selected by FOF($l = 0.1$) and FOF($l = 0.2$) respectively. The solid curve shows the results for haloes selected by an algorithm which mimics the definition of haloes in the Press-Schechter formalism (see text). The predictions of our analytic model (equation 23) for the same model parameters are shown as crosses. The value of $N_\mathrm{h}$ in each panel gives

the number of haloes in the simulation box.

**Figure 6.** The correlation length $r_0$, defined as the scale where the two-point correlation function is unity, as a function of the mean separation of clusters $d$ (defined from the mean number density of clusters $n$ by $d \equiv n^{-1/3}$). The dashed curve shows the result of Bahcall and Cen (1992) for a CDM model with $\sigma_8 = 0.75$. The solid curve shows the result of our analytic model for the same case. For comparison, we also show the observational data, as reported by Bahcall and Cen (1992).

**Figure 7.** Lagrangian average cross-correlations between haloes and mass as given by equation (18). Since the initial density field has been linearly extrapolated to $z = 0$, these cross-correlations can be smaller than $-1$. Results are shown for CDM models with $\sigma_8 = 1$ (7a) and $\sigma_8 = 0.5$ (7b), and for dark haloes with circular velocity of $v_c = 50$ (dotted), 100 (short-dashed), 200 (long-dashed), 400 (dotted-short-dashed), 800 (dotted-long-dashed), and 1600 km s$^{-1}$ (short-dashed-long-dashed curve). For comparison, average mass autocorrelation functions are shown as solid curves.

**Figure 8.** Eulerian average cross-correlations between haloes and mass as given by equation (22). Results are shown for CDM models with $\sigma_8 = 1$ (8a) and $\sigma_8 = 0.5$ (8b), and for dark haloes with circular velocities $v_c = 50$ (dotted), 100 (short-dashed), 200 (long-dashed), 400 (dotted-short-dashed), 800 (dotted-long-dashed), and 1600 km s$^{-1}$ (short-dashed-long-dashed curve), respectively. Average mass autocorrelation functions are shown as solid curves.

**Figure 9.** Lagrangian average autocorrelation functions of haloes calculated from equation (21). Results are shown for CDM models with $\sigma_8 = 1$ (9a) and $\sigma_8 = 0.5$ (9b), and for dark haloes with circular velocities $v_c = 50$ (dotted), 100 (short-dashed), 200 (long-dashed), 400 (dotted-short-dashed), 800 (dotted-long-dashed), and 1600 km s$^{-1}$ (short-dashed-long-dashed curve). Average mass autocorrelation functions are shown as solid curves.

**Figure 10.** Eulerian average autocorrelation functions of haloes given by equation (23). Results are shown for CDM models with $\sigma_8 = 1$ (10a) and $\sigma_8 = 0.5$ (10b), and for dark haloes with circular velocities $v_c = 50$ (dotted), 100 (short-dashed), 200 (long-dashed), 400 (dotted-short-dashed), 800 (dotted-long-dashed), and 1600 km s$^{-1}$ (short-dashed-long-dashed curve). Average mass autocorrelation functions are shown as solid curves.

**Figure 11.** Present-day average autocorrelation functions for haloes with the same circular velocity $v_c = 200 \, \mathrm{km \, s^{-1}}$, but identified at different redshifts. These correlation functions are calculated from equation (23) for $z_1 = z_2 = z$, but with the density perturbation $(\delta_0, R_0)$ evolved to $z = 0$. Results are shown for CDM models with $\sigma_8 = 1$ (11a) and $\sigma_8 = 0.5$ (11b), and for identification redshift $z = 0$ (dotted), 1 (short-dashed), 2 (long-dashed), 3 (dotted-short-dashed), and 4 (dotted-long-dashed). The present-day average mass autocorrelation functions are shown as solid curves.

**Figure 12.** Halo overdensity $\delta_\mathrm{h}$ as a function of mass overdensity $\delta_\mathrm{m}$ (equation 32) within a sphere with Eulerian radius $R = 20$ Mpc, (a) for haloes identified at $z = 0$ and with $v_c = 50, 100, 200, 400, 800,$ and $1600 \, \mathrm{km \, s^{-1}}$ (from flat to steep curves), and (b) for $v_c^*$ haloes at $z = 0, 1, 2, 3,$ and 4 (from the most curved to the least curved curves). Solid curves show results for $\sigma_8 = 1$; dashed ones show those for $\sigma_8 = 0.5$. The dotted lines show $\delta_\mathrm{h} = \delta_\mathrm{m}$.